\documentstyle[12pt]{article}

\font\blackboard=msbm10 at 12pt
\font\blackboards=msbm7
\font\blackboardss=msbm5
\newfam\black
\textfont\black=\blackboard
\scriptfont\black=\blackboards
\scriptscriptfont\black=\blackboardss

\newcommand{\ba}{\begin{array}}
\newcommand{\ea}{\end{array}}
\newcommand{\be}{\begin{equation}}
\newcommand{\ee}{\end{equation}}
\newcommand{\bea}{\begin{eqnarray}}
\newcommand{\eea}{\end{eqnarray}}
\newcommand{\beas}{\begin{eqnarray*}}
\newcommand{\eeas}{\end{eqnarray*}}

\def\half{{1 \over 2}}
\def\identity{{\rlap{1} \hskip 1.6pt \hbox{1}}}

\def\laplace{{\kern1pt\vbox{\hrule height 1.2pt\hbox{\vrule width 1.2pt\hskip
  3pt\vbox{\vskip 6pt}\hskip 3pt\vrule width 0.6pt}\hrule height 0.6pt}
  \kern1pt}}
\def\scriptlap{{\kern1pt\vbox{\hrule height 0.8pt\hbox{\vrule width 0.8pt
  \hskip2pt\vbox{\vskip 4pt}\hskip 2pt\vrule width 0.4pt}\hrule height 0.4pt}
  \kern1pt}}

\def\roughly#1{\raise.3ex\hbox{$#1$\kern-.75em\lower1ex\hbox{$\sim$}}}

\def\tr{{\rm Tr} \,}
\def\str{{\rm STr} \,}

\def\hx{{\hat{\bf X}}}
\def\hf{{\hat{F}}}
\def\hatt{{\hat{\cal T}}}
\def\hj{{\hat{\cal J}}}
\def\hm{{\hat{\cal M}}}
\def\sx{{\tilde{\bf X}}}
\def\sf{{\tilde{F}}}
\def\st{{\tilde{\cal T}}}
\def\sj{{\tilde{\cal J}}}
\def\sm{{\tilde{\cal M}}}
\def\itt{{\cal  T}}
\def\ijj{{\cal J}}
\def\imm{{\cal M}}

\def\dx{{\dot{\bf X}}}

\newcommand{\NP}{{\em Nucl.\ Phys.\ }}

\newcommand{\PRL}{{\em Phys.\ Rev.\ Lett.\ }}

\newcommand{\gone}[1]{}
\begin{document}
\pagestyle{plain}
\setcounter{page}{1}

\baselineskip16pt

\begin{titlepage}

\begin{flushright}
IASSNS-HEP-97/141\\
PUPT-1751\\
hep-th/9712185
\end{flushright}
\vspace{16 mm}

\begin{center}
{\Large \bf Linearized supergravity from Matrix theory}

\vspace{3mm}

\end{center}

\vspace{8 mm}

\begin{center}

Daniel Kabat${}^1$ and Washington Taylor IV${}^2$

\vspace{3mm}

${}^1${\small \sl School of Natural Sciences} \\
{\small \sl Institute for Advanced Study} \\
{\small \sl Olden Lane} \\
{\small \sl Princeton, New Jersey 08540, U.S.A.} \\
{\small \tt kabat@ias.edu}

\vspace{3mm}

${}^2${\small \sl Department of Physics} \\
{\small \sl Joseph Henry Laboratories} \\
{\small \sl Princeton University} \\
{\small \sl Princeton, New Jersey 08544, U.S.A.} \\
{\small \tt wati@princeton.edu}

\end{center}

\vspace{8mm}

\begin{abstract}

We show that the linearized supergravity potential between two objects
arising from the exchange of quanta with zero longitudinal momentum is
reproduced to all orders in $1/r$ by terms in the one-loop Matrix
theory potential.  The essential ingredient in the proof is the
identification of the Matrix theory quantities corresponding to
moments of the stress tensor and membrane current.  We also point out
that finite-$N$ Matrix theory violates the equivalence principle.

\end{abstract}

\vspace{8mm}
\begin{flushleft}
December 1997
\end{flushleft}
\end{titlepage}
\newpage

\section{Introduction}
\label{sec:Intro}

It has by now been firmly established that the matrix quantum
mechanics \cite{Claudson-Halpern,Flume,brr} appearing in the BFSS
conjecture \cite{BFSS} contains within it a remarkable structure which
replicates many features of 11-dimensional supergravity.  As
emphasized in the lectures of Susskind \cite{Susskind-review},
however, there is as yet no explicit way of deriving supergravity
directly from Matrix theory.  In this paper we make some progress in
this direction by deriving the linearized form of supergravity from a
subset of the terms in the one-loop Matrix theory potential.

Supergravity has two propagating bose fields: the metric and the
3-form.  The effective potential between an arbitrary pair of
classical objects is in principle given by summing all Feynman
diagrams connecting the objects.  When this summation is truncated to
include only single-particle exchange diagrams, the resulting
potential corresponds to that of linearized supergravity.  For a pair
of objects of finite but nonzero size, this potential can be expanded
in the inverse separation, giving a power series in $1/r$, where a
term of order $1/r^{7 + k + l}$ appears for each $k$ and $l$ which is
proportional to the product of the $k$th and $l$th moments of the
source tensors of the two objects.  In this paper we prove that this
infinite series of terms corresponds precisely to a series of terms in
the one-loop Matrix theory potential.  The leading term in this series
is the $1/r^7$ term proportional to the product of the integrated
source tensors.  This term has been shown to be reproduced correctly
by Matrix theory in numerous specific cases
\cite{BFSS,DKPS,Aharony-Berkooz,Lifschytz-Mathur,Lifschytz-46,bc,ChepTseyI,MaldacenaI,Vakkuri-Kraus,Gopakumar-Ramgoolam,ChepelevTseytlin,MaldacenaII,Esko-Per2};
however, to date there has been no general understanding of the
structure responsible for this agreement.

It should be emphasized that the higher order terms we consider in
this paper are terms of the form $F^4 X^n$, which were considered
previously in \cite{Mark-Wati}.  These should not be confused with the
terms of higher order in $F$ in the one-loop potential or the higher
loop terms which have been discussed in the context of graviton
scattering
\cite{Vijay-Finn2,Becker-Becker,ggr,bbpt,Chepelev-Tseytlin2}.

The structure of this paper is as follows: In Section 2 we recall the
general form of the leading term in the one-loop Matrix theory
potential between an arbitrary pair of objects, and show that this
term can be rewritten in a fashion which is highly suggestive of
supergravity interactions.  In Section 3 we derive the leading term in
the linearized supergravity potential arising from the exchange of
quanta with no longitudinal momentum.  In Section 4 we observe that
the Matrix theory and supergravity potentials agree precisely if
certain Matrix expressions are identified as the stress tensor and
currents of a Matrix theory object.  We show that in the large-$N$
limit this identification corresponds with known results for the
graviton, membrane, and longitudinal 5-brane.  Section 5 demonstrates
that the subleading terms in the linearized supergravity potential
arising from higher moments of the sources are also correctly
reproduced to all orders in Matrix theory.  Section 6 concludes with a
discussion of the relationship of supergravity to finite--$N$ Matrix
theory.

\section{Matrix theory potential}
\label{sec:matrix-potential}

We will consider the leading term in the long-distance expansion of
the Matrix theory potential between an arbitrary pair of localized
semiclassical states.  Two objects described by matrices $\hat{{\bf
X}}(t)$ and $\tilde{\bf X}(t)$ of sizes $\hat{N}\times\hat{N}$ and
$\tilde{N}\times\tilde{N}$, respectively, correspond to a background
in the Matrix gauge theory
\[
\left\langle {\bf X}^i(t) \right\rangle = \left(
\begin{array}{cc}
\hat{\bf X}^i(t) & 0 \\
0 & \tilde{\bf X}^i(t)
\end{array}
\right) \,.
\]
Integrating out the off-diagonal matrix blocks at one loop gives rise
to the leading long-distance potential \cite{Chepelev-Tseytlin, Dan-Wati}
\be
\label{eq:MatrixPotential}
V_{{\rm matrix}} = -\frac{5}{128\;r^7} \str {\cal F}
\ee
where $r$ is the separation distance between the objects, ${\cal F}$
is defined by
\bea
{\cal F} & = & 
      24 \; F_{0i} F_{0i} F_{0j} F_{0j}
    + 24 \;F_{0i} F_{0i} F_{jk} F_{jk}
    + 96 \;F_{0i} F_{0j} F_{ik} F_{kj} \\
& & \qquad \qquad + 24 \; F_{ij} F_{jk} F_{kl} F_{li}
    -  6 \;F_{ij} F_{ij} F_{kl} F_{kl} \nonumber
\label{eq:W}
\eea
and $\str$ is a symmetrized trace, the average over all
orderings in the product of $F$'s.
The components of $F$ are defined by
\[
F_{0i} = - F_{i0} = \partial_t K_i, \quad F_{ij} = i \left[K_i,K_j\right]
\]
where
\[
K_i = \hx_i \otimes \identity_{\tilde{N}\times\tilde{N}}
- \identity_{\hat{N}\times\hat{N}}\otimes \sx_i^T\,.
\]
The potential can be expanded in terms of the field
strengths of the individual objects
\[
\hat{F}_{0i} = - \hat{F}_{i0} = \partial_t \hat{{\bf X}}_i,
\;\;\;\;\;\hat{F}_{ij} = i [\hat{{\bf X}}_i, \hat{{\bf X}}_j]
\]
\[
\tilde{F}_{0i} = - \tilde{F}_{i0} = \partial_t \tilde{\bf X}_i,
\;\;\;\;\; \tilde{F}_{ij} = i [\tilde{\bf X}_i, \tilde{\bf X}_j].
\]
(In the sequel all variables with a hat (tilde) indicate quantities
which depend only on the matrices $\hat{{\bf X}} (\tilde{\bf X})$).
By decomposing the Matrix potential (\ref{eq:MatrixPotential}) in
terms of $\hf$ and $\sf$, it can be re-expressed as\footnote{Notation:
$I,J = 0,\ldots,10$ are spacetime indices with metric $\eta_{IJ} =
\left(-+\cdots+\right)$, $x^\pm = {1 \over \sqrt{2}} \left(x^0 \pm
x^{10}\right)$ with $x^- \approx x^- + 2 \pi R$, and $i,j =
1,\ldots,9$ are transverse spatial indices.} 
\bea
\label{eq:MatrixPotentialII}
V_{\rm matrix} & = & V_{\rm gravity} + V_{\rm electric} + V_{\rm magnetic} \\
V_{\rm gravity} & = & - {15 R^2 \over 4 r^7} \left( \hatt^{IJ} \st_{IJ}
- {1 \over 9} \hatt^I{}_I \st^J{}_J\right) 
\label{eq:v-gravity} \\
V_{\rm electric} & = & - {45 R^2 \over r^7} \hj^{IJK} \sj_{IJK} 
\label{eq:v-electric} \\
V_{\rm magnetic} & = & - {45 R^2\over r^7} \hm^{+-ijkl} \sm^{-+ijkl} 
\label{eq:v-magnetic}
\eea
The quantities appearing in this decomposition are defined as follows.
(One can verify these relations simply by collecting all terms on both sides with a
definite tensor structure in either $\hf$ or $\sf$.)  

$\itt^{IJ}$ is a
symmetric tensor with components
\bea
\label{eq:MatrixT}
\itt^{--} & = & {1 \over R} \; \str \frac{{\cal F}}{96}  \\
\itt^{-i} & = & {1 \over R} \;\str \left(\half \dx^i \dx^j \dx^j +
{1 \over 4} \dx^i F^{jk} F^{jk} 
                  + F^{ij} F^{jk} \dx^k \right) \nonumber \\
\itt^{+-} & = & {1 \over R} \;\str \left(\half \dx^i \dx^i + {1
\over 4} F^{ij} F^{ij} \right) \nonumber \\ 
\itt^{ij} & = & {1 \over R}  \;\str \left( \dx^i \dx^j + F^{ik}
F^{kj} \right) \nonumber \\ 
\itt^{+i} & = & {1 \over R} \;\str \dx^i \nonumber \\
\itt^{++} & = & {N \over R} \nonumber
\eea
$\ijj^{IJK}$ is a totally antisymmetric tensor with components
\bea
\label{eq:MatrixJ}
\ijj^{-ij} & = & {1 \over 6 R} \str \left( \dx^i \dx^k F^{kj} -
\dx^j \dx^k F^{ki} - \half \dx^k \dx^k F^{ij} \right.\\ 
& & \qquad \qquad  \left. + {1 \over 4} F^{ij} F^{kl} F^{kl} +F^{ik}
F^{kl} F^{lj} \right) \nonumber \\ 
\ijj^{+-i} & = & {1 \over 6 R} \str \left( F^{ij} \dx^j \right) \nonumber \\
\ijj^{ijk} & = & - {1 \over 6 R} \str \left( \dx^i F^{jk} + \dx^j F^{ki} + \dx^k F^{ij} \right) \nonumber \\
\ijj^{+ij} & = & - {1 \over 6 R} \str F^{ij} \nonumber
\eea
Note that we retain some quantities -- in particular $\ijj^{+-i}$ and
$\ijj^{+ij}$ -- which vanish at finite $N$ (by the Gauss constraint and
antisymmetry of $F^{ij}$, respectively).  These terms
represent BPS charges (for longitudinal and transverse membranes)
which are only present in the large $N$ limit \cite{bss}.  We define higher
moments of these terms in Section \ref{sec:higher} which can be
non-vanishing at finite $N$.

$\imm^{IJKLMN}$ is a totally antisymmetric tensor with
\be
\label{eq:MatrixM}
 \imm^{+-ijkl} = {1 \over 12 R} \str \left(F^{ij} F^{kl} + F^{ik}
F^{lj} + F^{il} F^{jk}\right)\,.
\ee
At finite $N$ this vanishes by the Jacobi identity, but we shall
retain it because it represents the charge of a longitudinal 5-brane.
The other components of ${\cal M}^{IJKLMN}$ do not appear in the Matrix
potential, and we are unable to determine expressions for them.  This
is closely related to the difficulty of constructing a transverse
5-brane in Matrix theory.

We have thus expressed the leading term in the one-loop Matrix theory
potential in an algebraic form which is highly suggestive of
supergravity.  This potential is a time-dependent instantaneous
interaction potential which depends upon various tensor expressions in
the Matrix variables describing the two objects.  We now compare this
with the form of the linearized supergravity potential between an
arbitrary pair of objects.

\section{Supergravity potential}
\label{sec:supergravity-potential}

We begin with a few general remarks on the supergravity potential in
light-front coordinates.  According to the BFSS conjecture
\cite{BFSS}, the leading term in the one-loop matrix theory potential
should agree with the leading long-distance supergravity interaction
between two objects when no longitudinal momentum is transfered.
Thus, it should be compared to the interaction expected from
linearized supergravity, in which the exchanged quantum (metric or
3-form) has zero longitudinal momentum.  In light-front coordinates,
this leads to a rather peculiar propagator, as emphasized in
\cite{Hellerman-Polchinski}.  For example, the scalar Green's function
is
\[
\laplace{}^{-1}(x) = {1 \over 2 \pi R} \sum_n \int{dk^- d^{\,9}k_\perp \over (2 \pi)^{10}}
{e^{-i{n \over R} x^- - i k^- x^+ + i k_\perp \cdot x_\perp} \over
2 {n \over R} k^- - k_\perp^2}
\]
Keeping just the $n=0$ term in the sum leads to the propagator at zero longitudinal momentum
\be
\label{eq:propagator}
\laplace{}^{-1}(x-y) = {1 \over 2 \pi R} \delta(x^+ - y^+) { - 15
\over 32 \pi^4 \vert x_\perp-y_\perp\vert^7 } \ee 
Note that the exchange of quanta with zero longitudinal momentum gives
rise to interactions that are instantaneous in light-front time,
precisely the type of instantaneous interactions that we find at one
loop in Matrix theory.  Such action-at-a-distance potentials are
allowed by the Galilean invariance manifest in light-front
quantization.

An object with internal dynamics (such as an oscillating
membrane) gives rise to gravitational radiation, which appears as
time-dependent fluctuations in the metric tensor seen by a distant
observer.  An observer using light-front coordinates will see part of
this radiation as an instantaneous potential, even though the
radiation respects causality.  This instantaneous potential carries no energy
away from the radiating object; to see a loss of energy due to
outgoing radiation it would be necessary to calculate the rate of
emission of gravitons with non-zero longitudinal momentum.

We now calculate the light-front supergravity interaction potential
at leading order in $1/r$.
We  take the bulk action for
11-dimensional supergravity to be\footnote{Conventions: The Planck length is
defined by $2 \kappa^2 = (2 \pi)^8 l_P^9$.  We adopt units in which $2
\pi l_P^3 = R$.  The membrane tension is $T_2 = {1 \over (2 \pi)^2
l_P^3}$, and $F_{IJKL} \equiv {1 \over 6} \left(\partial_I C_{JKL} \pm
\hbox{\rm 23 terms}\right)$.}
\[
S_{\rm SUGRA} = - {1 \over 2 \kappa^2} \int d^{11}x \sqrt{-g} \left(R + {1 \over 24} F_{IJKL} F^{IJKL} + \cdots\right)
\]
where $F$ is the field strength of the 3-form potential $C$.
Integrating out the metric and 3-form in the linearized approximation
gives rise to the effective action
\[
S = \int d^{11} x d^{11} y \, \left( - {1 \over 8} T_{IJ}(x) D_{\rm
graviton}^{IJ,KL}(x-y) T_{KL}(y)  - {1 \over 2} J_{IJK}(x) D_{\rm
3-form}^{IJK,LMN}(x-y) J_{LMN}(y) \right)
\]
where the gauge-fixed graviton and 3-form propagators are
\beas
D_{\rm graviton}^{IJ,KL} & = & 2 \kappa^2 \left(\eta^{IK} \eta^{JL} + \eta^{IL} \eta^{JK} - {2 \over 9}
\eta^{IJ} \eta^{KL} \right) \laplace{}^{-1}(x-y) \\
D_{\rm 3-form}^{IJK,LMN} & = & 2 \kappa^2 \left(\eta^{IL} \eta^{JM} \eta^{KN} \pm \hbox{\rm 5 terms}\right)
\laplace{}^{-1}(x-y)
\eeas
and where $T^{IJ}$ and $J^{IJK}$ are the stress tensor and `electric'
3-form current.
To compare with Matrix theory, we use the zero longitudinal momentum propagator (\ref{eq:propagator})
to find the potential between two objects arising from graviton and electric 3-form exchange:
\beas
V_{\rm sugra} &=& V_{\rm gravity} + V_{\rm electric} \\
V_{\rm gravity} &=& - {15 R^2 \over 4 r^7} \left(\hatt^{IJ} \st_{IJ} - {1 \over 9} \hatt^I{}_I
\st^J{}_J\right) \\
V_{\rm electric} &=& - {45 R^2 \over r^7} \hj^{IJK} \sj_{IJK}
\eeas
where we define, for example, $\itt^{IJ} \equiv \int dx^- d^{\, 9}
x_\perp T^{IJ}(x)$. 

The 3-form field also has `magnetic' couplings to the M-theory
5-brane.  Although it is not known how to describe a theory containing
both electric and magnetic sources in terms of an action principle,
in the linearized theory we can simply consider interactions between
magnetic sources to be mediated by a quantum of the 6-form field dual
to $C$.  This leads to an additional term in the supergravity
potential
\[
V_{\rm magnetic} = - {3 R^2 \over 2 r^7} \hm^{IJKLMN}
\sm_{IJKLMN}. 
\]
where ${\cal M}^{IJKLMN}$ is the `electric' current for the 6-form
field.

\section{Correspondence of potentials}
\label{sec:correspondence}

The one-loop Matrix potential (\ref{eq:MatrixPotentialII}) clearly has
a structure identical to the linearized supergravity potential derived
in the last section.  At a formal level, this shows that if we {\em
define} the stress tensor and currents of objects in Matrix theory to
be given by (\ref{eq:MatrixT}), (\ref{eq:MatrixJ}) and
(\ref{eq:MatrixM}) then there is a precise correspondence between the
Matrix theory and supergravity potentials between any pair of objects.
This formal correspondence is valid even at finite $N$.  However, to
verify that this equivalence is physically meaningful, we must check
that the Matrix theory stress tensor and currents we have defined
correspond in a sensible way with those of supergravity.  In this
section we show that this correspondence does indeed hold in the
large-$N$ limit; finite-$N$ effects are discussed in Section 6.

To show that our definitions of the Matrix theory stress tensor and
currents are sensible, we now proceed to show that they correspond with
the definitions expected from supergravity for three classes of
semiclassical states: gravitons, membranes, and longitudinal 5-branes.

\subsection{Gravitons}

We begin by discussing graviton states.  A classical graviton 
with transverse coordinates $x^i(t)$ and $N$ units of longitudinal
momentum is described in Matrix theory by matrices ${\bf X}^i = x^i
\identity_{N
\times N}$.  This is sufficient for our purposes because at the
one-loop level the graviton wavefunction does not enter: all we need
is a background which satisfies the classical Matrix equations of
motion.  Plugging this background into the Matrix definitions
(\ref{eq:MatrixT}), (\ref{eq:MatrixJ}) and (\ref{eq:MatrixM}) one finds
that, as expected, the 3-form and 6-form currents vanish, while the
stress tensor is given by
\beas
 \itt^{++} = {N \over R} {\rm \hspace{0.55in}} \itt^{+i}  &=   &
\frac{N}{R}  \dot{x}^i
{\rm \hspace{0.55in}}  \itt^{ij} = {N \over R} \dot{x}^i \dot{x}^j \\
 \itt^{+-} = {N \over 2R} \dot{x}^2
\hspace{0.55in} \itt^{-i}  &=  &
\frac{N}{2 R}  \dot{x}^i \dot{x}^2 
\hspace{0.45in}   \itt^{--} = {N \over 4R} \left(\dot{x}^2\right)^2
\eeas
This matches the supergravity result: the classical expression for the stress tensor of a point particle is
\[
\int dx^- d^{\,9} x_\perp T^{IJ}(x) = {p^I p^J \over p^+}
\]
where for a massless particle $p^+ = N/R$, $p^i = p^+ \dot{x}^i$, and $p^- = p_\perp^2 / 2 p^+$.

\subsection{Membranes}

The bosonic part of the supermembrane action is
\be
\label{eq:MembraneAction}
S_{\rm membrane} = \int d^3 \xi \left\lbrace - T_2 \sqrt{-{\rm det}
\partial_\alpha X^I \partial_\beta X^J g_{IJ}} 
+ {\sqrt{2}\,T_2 \over 3!}\epsilon^{\alpha\beta\gamma}\partial_\alpha
X^I \partial_\beta X^J \partial_\gamma X^K 
C_{IJK} \right\rbrace\, .
\ee
The resulting stress tensor and 3-form current
are given by
\bea
\label{eq:SugraMembraneCurrents}
T^{IJ}(x) &=& - T_2 \int d^3 \xi \, \delta^{11}(x - X(\xi)) \,
\sqrt{-g} g^{\alpha \beta} 
\partial_\alpha X^I \partial_\beta X^J \\
J^{IJK}(x) &=& T_2 \int d^3 \xi \, \delta^{11}(x - X(\xi)) \, {1 \over
3!} \epsilon^{\alpha\beta\gamma} 
\partial_\alpha X^I \partial_\beta X^J \partial_\gamma X^K
\eea
where $g_{\alpha\beta} = \partial_\alpha X^I \partial_\beta X^J
g_{IJ}$ is the induced metric on the membrane. 

There is a remarkable connection between supermembrane theory in
light-front gauge and Matrix theory
\cite{Goldstone-membrane,Hoppe-membrane,bst2,dhn}.  In light-front gauge,
with coordinates $\xi^\alpha = (t,\sigma^a)$ on the membrane worldvolume,
the dynamics may be expressed in terms of a Poisson bracket $\{f,g\} \equiv \epsilon^{ab} \partial_a f \partial_b g$.
The fields $X^+$, $X^-$ are constrained by
\bea
X^+ &=& t  \nonumber\\
\noalign{\vskip 0.2 cm}\label{constraints}
\dot{X}^-  &=& \half \dot{X}^i \dot{X}^i + {1 \over N^2}
\{X^i,X^j\}\{X^i,X^j\} \nonumber\\
\noalign{\vskip 0.2 cm}
\partial_a X^- &=& \dot{X}^i \partial_a X^i \nonumber
\eea
hence the integrated sources can be expressed as
\bea
\int dx^- d^{\,9}x_\perp T^{IJ} &=&  T_2 N \int d^2 \sigma \left(
 \half \dot{X}^I \dot{X}^J
- {2 \over N^2} \gamma \gamma^{ab} \partial_a X^I \partial_b X^J
\right) 
\label{eq:membrane-T}\\
\int dx^- d^{\,9}x_\perp J^{IJK} &=& T_2 \int d^2 \sigma \left(
 {1 \over 3!} \epsilon^{\alpha\beta\gamma}
\partial_\alpha X^I \partial_\beta X^J \partial_\gamma X^K \right)
\label{eq:membrane-J}
\eea
where $\gamma_{ab} \equiv g_{ab}$.
Expressions in terms of the light-front membrane fields can be translated into
their counterparts in Matrix theory by replacing Poisson
brackets with matrix commutators, and integrals with traces:
\be
\label{eq:MatMemCorrespondence}
\{\cdot,\cdot\} \leftrightarrow  {-i N \over 2} [\cdot,\cdot] \hskip 1.0 cm
{N \over 4 \pi} \int d^2 \sigma \leftrightarrow \tr \,.
\ee
Using this correspondence and the constraints on $X^+$ and $X^-$ it is
straightforward to show that the expressions
(\ref{eq:membrane-T}) and (\ref{eq:membrane-J}) agree with
the Matrix theory definitions (\ref{eq:MatrixT}) and
(\ref{eq:MatrixJ}), and that the 6-form current (\ref{eq:MatrixM})
vanishes; the identity $\{X^i,X^k\} \{X^j,X^k\} = \gamma \gamma^{ab} \partial_a X^i \partial_b X^j$
is useful in proving these relations.  

There is only one subtlety in this correspondence.  Taking the membrane expression
(\ref{eq:membrane-J}) for the component $J^{-ij}$ and translating into
Matrix theory language through (\ref{eq:MatMemCorrespondence}) we find
that the resulting current is not precisely equal to
(\ref{eq:MatrixJ}).  We find that the Matrix theory current
$\ijj^{-ij}$ contains an extra term of the form
\[
\frac{1}{6 R}
  \str \left({1 \over 2} F^{ij} F^{kl} F^{kl} + F^{ik} F^{kl} F^{lj} \right).
\]
Translating this back into membrane language, we see that it vanishes
identically.  Thus, this term in the Matrix theory expression is
unaffected by a single membrane source.  We have not found a physical
interpretation for this term.

One final note of interest with regard to the membrane is that the
correspondence (\ref{eq:MatMemCorrespondence}) is only exact
in the large-$N$ limit.  Thus, the Matrix theory stress tensor and
currents are only identified with the membrane stress tensor and
currents up to $1/N$ corrections.  We will return to this issue in the
final section.

\subsection{Longitudinal 5-branes}

To complete our comparison of the Matrix theory expressions for the
stress tensor and source terms with supergravity, we consider the only
other extended object which has been found in Matrix theory: the
longitudinal 5-brane.  Flat L5-branes were found in \cite{grt,bss},
where it was shown that the L5-brane charge is proportional to
$\imm^{+-ijkl}$.  Finite size L5-branes with the transverse geometry
of a 4-sphere were constructed in \cite{clt}.  We will now discuss the
matrix expressions for the stress tensor and the sources for an
L5-brane sphere of radius $r$; to get the results for a flat L5-brane
one can take $r \rightarrow \infty$ and focus on a region of the
sphere which is locally flat.

We assume that our L5-brane sphere is extended in directions $1-5$ and
has radius $r$, with $ \dot{r} = 0$ for simplicity.  For $n$
superimposed L5-brane spheres we have $N = (n + 1) (n +2) (n + 3)/6$.
Using the explicit expressions for this object from \cite{clt}, it is
easy to calculate the various tensors.  As for all Matrix theory
objects, we have $\itt^{+ +} = N/R$ and $\itt^{+ -}= E$ where $E$ is
the light-front (Matrix theory) energy.  The only other nonzero
components of the stress tensor are $\itt^{ij}$ and $\itt^{--}$.  The
first of these is equal to $-4E \delta^{ij}/5$ in the five directions
in which the brane is extended.  The component $\itt^{--}$ vanishes at
leading order in $1/N$, as discussed in \cite{clt}.  This is in accord
with the fact that a static 0-brane (graviton) should feel no force
from a 4-brane (L5-brane).  However, there are contributions at lower
order which make this term non-vanishing at finite $N$.  All
components of $\ijj$ vanish, as one would expect since the L5-brane
should not be an electric source for the 3-form field.  Finally,
although the net L5-brane charge is zero and the trace giving the
source term $\imm^{+-ijkl}$ vanishes, the higher moments of this term
(discussed in the following section) are non-vanishing, and can be
seen to give the correct values (up to $1/N$ corrections) from the
fact that the L5-brane charge locally has the correct values, as
discussed in \cite{clt}.

Thus, we see that the stress tensor and currents defined in
(\ref{eq:MatrixT}), (\ref{eq:MatrixJ}) and (\ref{eq:MatrixM}) agree with
the values we expect from the correspondence between the known
Matrix theory objects and their supergravity counterparts.  It is
interesting that we have seen no sign of other components of the
6-form current ${\cal M}$.  This is presumably related to the absence
of transverse 5-brane charge in the superalgebra \cite{bss} and is a
reflection of the well-known difficulty of constructing a transverse
5-brane in Matrix theory.

\section{Higher order terms}
\label{sec:higher}

In the supergravity potential between two extended objects, there are
terms arising from single particle exchange which appear at higher
orders in $1/r$.  These terms arise from higher moments of the
stress-energy and 3-form source tensors of the two objects.  In this
section, we show that these terms are precisely replicated by an
infinite series of terms in the one-loop Matrix theory potential.
This result was shown in the special case of the membrane-graviton
potential in \cite{Mark-Wati}; here we generalize to the case of two
extended objects and general source terms.  The terms we consider in
the one-loop Matrix theory potential were calculated using the
quasi-static approximation.  It is known \cite{Tafjord-Periwal} that
the results of such a calculation can differ from those calculated using the
phase-shift method; however, the quasi-static approximation should be
correct for the terms of the form $F^4 X^n$ which we consider here.

In the quasi-static approach to calculating the matrix theory
potential described in \cite{Dan-Wati}, there is an infinite
series of terms which are of order $F^4$ but which have extra factors
of $K/r$ inserted into the trace.
This series of terms was calculated in the membrane-graviton case in
\cite{Mark-Wati}.  To generalize to the case of two extended objects
we will use the convention that $\hat{{\bf X}}$ and $\tilde{\bf X}$
describe the two objects in their respective centers of mass at time
$t$, while $r_i$ denotes the separation at this time.
The full series of terms contributing to the potential at order $F^4$
can then be expressed as
\begin{eqnarray}
V_{{\rm matrix}}
& = & \sum_{n = 0}^{\infty} \sum_{k \leq n/2}
\frac{ (-1)^{k + 1}\; (5 + 2 n-2k)!!}{3 \cdot 2^{7 + k} \; k! \; (n-2k)!
\; \; r^{7 +2n-2k}} \; {\rm STr}\; \left(
(r \cdot K)^{n-2k} K^{2k} {\cal F} \right).\label{eq:matrix-many}
\end{eqnarray}
The trace is symmetrized with respect to the matrices $r \cdot K, K^2$
and $F$.  Just as the leading term was separated in
(\ref{eq:MatrixPotentialII}) into terms having different forms of Lorentz
index contraction between the two traces, the same decomposition can
be performed in all the subleading terms.  For example, at the first
subleading order we have
\begin{equation}
-\frac{35}{128 \;r^9} 
{\rm STr}\; \left[ (r \cdot\hat{{\bf X}} -r \cdot \tilde{\bf X}^T)
{\cal
F} \right] =
\frac{245 R^2}{6 \; r^9} 
r_i \left(\hatt^{+-(i)} \st^{+-} -\hatt^{+-}
\st^{+-(i)} \right) + \cdots
\label{eq:decomposition-1}
\end{equation}
where we define the moments of $\itt^{+-}$ through
\begin{equation}
\itt^{+-(i_1 i_2 \cdots i_n)} = {1 \over R} {\rm STr}\; \left[ \left(\half \dx^i \dx^i + {1
\over 4} F^{ij} F^{ij}  \right) {\bf X}^{i_1} {\bf X}^{i_2}\cdots
{\bf X}^{i_n} \right]\ .
\label{eq:moments}
\end{equation}
The moments of all of the components of the stress tensor and currents
can be defined in an analogous fashion, and all the symmetrized
moments of $\str {\cal F}$ can be shown to decompose in an analogous
fashion to (\ref{eq:MatrixPotentialII}).  Note that there is a
subtlety here when a pair of indices in the moment
are contracted; in this case, the trace is symmetrized with respect to
the square ${\bf X}^2$ so that the separate ${\bf X}$'s cannot be
separated, as in (\ref{eq:matrix-many}).  This unusual property of the
moments is presumably related to the effects of noncommutative geometry at
finite $N$ discussed in the final section.

In order to compare this matrix theory calculation with supergravity,
we need to determine the effects of higher moments in the sources
associated with $\itt$, $\ijj$ and $\imm$.  Generalizing the argument of
\cite{Mark-Wati} to two-body systems, we have for the stress tensor
\begin{eqnarray}
V_{{\rm  gravity}}  &= &
\sum_{m  \leq n = 0}^{\infty}
-\frac{15 R^2}{4 \; r^7}  \left[
\frac{(-1)^{n-m}}{(n-m) ! m !}
\hatt^{I J(i_1i_2 \cdots i_{n-m})} \left(
 \eta_{I K} \eta_{J L}
-\frac{1}{9}  \eta_{I J} \eta_{K L} \right)
\st^{K L(j_1j_2 \cdots j_m)} \right. \nonumber\\
& &\hspace{1.2in} \left.\times
\partial_{i_1} \partial_{i_2} \cdots
\partial_{i_{n-m}}
\partial_{j_1} \partial_{j_2} \cdots
\partial_{j_m}
 (\frac{1}{r^7}) \right].
\label{eq:graviton-many}
\end{eqnarray}
where the moments of the stress tensor in the supergravity theory are
defined through
\[
\itt^{IJ (i_1 i_2 \cdots i_n)}  \equiv \int dx^- d^{\, 9}
x_\perp \left( T^{IJ}(x) x^{i_1} x^{i_2} \cdots x^{i_n} \right).
\]
It is a straightforward exercise in combinatorics to verify
that (\ref{eq:graviton-many}) reproduces the terms in
(\ref{eq:matrix-many}) associated with higher moments of the terms in
(\ref{eq:v-gravity}).  The proof of the analogous statement for the
higher moments of the source terms in (\ref{eq:v-electric}) and
(\ref{eq:v-magnetic}) follows in exactly the same fashion.

Thus, we have shown that all terms in the supergravity potential
arising from the propagation of a single graviton or 3-form quantum
between a pair of extended objects  are precisely matched by terms in
the one-loop matrix theory potential of the form $F^4 X^n$, where the
terms with an insertion of $X^n$ correspond to higher moments of the
source tensors associated with the extended objects.

\section{Finite $N$ and the equivalence principle}
\label{sec:FiniteN}

As was shown in section 2, the Matrix theory potential can be written
in a form (\ref{eq:MatrixPotentialII}) which is highly reminiscent of
supergravity.  The sources $\itt$, $\ijj$ and $\imm$ which appear in the
potential are traces of polynomials in the Matrix theory variables; if
we take these quantities to be the definitions of the
stress tensor and the 3-form and 6-form currents then the Matrix
potential is formally identical to the supergravity potential.

In the large-$N$ limit we showed that this definition is sensible: these
quantities agree with the corresponding supergravity expressions.  But the Matrix
expressions for the sources are perfectly well-defined at finite $N$.
Is finite-$N$ Matrix theory also related to supergravity?  A number of
recent arguments
\cite{Sen,Seiberg-DLCQ,dealwis-DLCQ,Balasubramanian-gl} in support of
the DLCQ conjecture \cite{Susskind-DLCQ} show that finite-$N$ Matrix
theory can be identified with DLCQ M-theory.  The question is then
whether DLCQ M-theory is described at low energies by DLCQ
supergravity.  But these two theories may well be different
\cite{BanksReview,Hellerman-Polchinski}, and indeed a number of explicit
calculations point to discrepancies
\cite{dos,ggr,Dine-Rajaraman,Douglas-Ooguri,Esko-Per-short}.

We now wish to point out that finite-$N$ Matrix theory, unlike
supergravity, does not obey the equivalence principle.  This may be
seen in the context of a simple example: consider the interaction of
two Matrix theory objects, a graviton initially
located at the origin with momentum
\[
\tilde{p}^+ = \tilde{N}/R \hskip 1.0 cm 
\tilde{p}^i = 0 \hskip 1.0 cm \tilde{p}^- = 0
\]
(described semiclassically by $\tilde{\bf X}^i = 0_{\tilde{N}
\times \tilde{N}}$), and a spherical membrane \cite{Dan-Wati}
initially located at transverse position $x^i$ with momentum
\[
p^+ = N/R \hskip 1.0 cm p^i = 0 \hskip 1.0 cm p^- = {8 r_0^4 \over R N^3} c_2 \,.
\]
The membrane is described semiclassically by
\[
{\bf X}^i = \left\lbrace
\begin{array}{ll}
x^i \identity_{N \times N} + {2 \over N} r_0 J^i & i=1,2,3 \\
x^i \identity_{N \times N} & {\rm otherwise}
\end{array}\right.
\]
where $r_0$ is the maximal radius of the sphere, $J^i$ are generators
of the $N$-dimensional representation of $SU(2)$, and $c_2 = {N^2 - 1
\over 4}$ is the corresponding quadratic Casimir.  We take $\tilde{N}
\gg N$, so that the graviton acts as a static gravitational source,
and study the resulting motion of the membrane.

The initial velocity of the membrane is
\[
\dot{x}^+ = 1 \hskip 1.0 cm \dot{x}^i = 0 \hskip 1.0 cm
\dot{x}^- = {p^- \over p^+} = 8 r_0^4 {c_2 \over N^4}\,.
\]
Consider changing $r_0$ while preserving the initial velocity.  That
is, consider scaling $r_0$ with $N$ according to $r_0^4 \sim {N^4
\over c_2}$.  The equivalence principle predicts that spheres with
different radii but identical initial velocities should fall at the
same rate in a uniform gravitational field: the initial acceleration
of the sphere should be independent of its mass, {\em i.e.} of $r_0$
and therefore $N$.

But this is not the case.  Using the one-loop Matrix potential one
finds that for large separation the initial acceleration is
\[
\ddot{x}^i = - {R \over N} \, {\partial V_{\rm matrix} \over \partial x^i} =
- 1680 R \tilde{N} {x^i \over \vert x \vert^9} \,\, 
{r_0^8 \over N^8} \left(c_2^2 - {1 \over 3} c_2 \right) \,.
\]
For large $N$, with the above scaling, the acceleration is indeed
independent of the mass.  But there are subleading terms of order
$1/N^2$ which violate the equivalence principle.\footnote{The
expressions for $\dot{x}^-$ and $\ddot{x}^i$ receive quantum
corrections, but these corrections are suppressed by a factor of
${l_P^3 \over N r_0^3}$ relative to the leading term and hence cannot
resolve this violation.}

It is important to emphasize that this violation of the equivalence
principle occurs in the leading long-distance interaction of
finite-$N$ Matrix theory.  The conclusion is that the leading
long-wavelength effective theory of DLCQ M-theory cannot be
Einstein-Hilbert gravity.  The breakdown of the equivalence principle
at finite $N$ is perhaps not surprising, since this principle follows
from commutative geometry.  Perhaps violation of the equivalence
principle in Matrix theory is associated with the noncommutative
nature of Matrix geometry: at finite $N$ there may be no limit in
which conventional geometry can be recovered.  It would be interesting
to find a quantitative explanation for this phenomenon in terms of
noncommutative geometry.

\section*{Acknowledgements}

We thank S.~de Alwis, T.~Banks, C.~Callan, J.~Castelino, M.~Gutperle,
I.~Klebanov, S.~Lee, G.~Lifschytz, V.~Periwal, N.~Seiberg,
L.~Thorlacius and M.~Van Raamsdonk for useful discussions.  The work
of DK is supported in part by the Department of Energy under contract
DE-FG02-90ER40542 and by a Hansmann Fellowship.  The work of WT is
supported in part by the National Science Foundation (NSF) under
contract PHY96-00258.

\bibliographystyle{plain}

\end{document}